\documentclass[10pt, journal, twocolumn]{IEEEtran}

\usepackage{epsfig}
\usepackage{amsmath,amsthm,amssymb,amsfonts}
\usepackage{bm}

\usepackage{color}

\usepackage[noend]{algpseudocode}
\usepackage{algorithmicx,algorithm}
\usepackage{hyperref}  

\begin{document}\sloppy

\def\x{{\mathbf x}}
\def\L{{\cal L}}

\title{A Practical Machine Learning Approach for Dynamic Stock Recommendation}
%
\author{Hongyang Yang$^{1}$, Xiao-Yang Liu$^{2}$, Qingwei Wu$^{2}$\\ 
$^1$AI4Finance Foundation\thanks{The AI4Finance Foundation (https://ai4finance.org) is a U.S.-registered 501(c)(3) nonprofit public charity focused on promoting open scientific research in financial AI, building open-source infrastructure, and supporting a global community of researchers through shared datasets, benchmarks, and educational programs.}\\
$^2$Dept. of Electrical Engineering, Columbia University\\
Email: contact@ai4finance.org
}

\maketitle

\begin{abstract}

Stock recommendation is vital to investment companies and investors. However, no single stock selection strategy will always win while analysts may not have enough time to check all S\&P 500 stocks (the Standard \& Poor's 500). In this paper, we propose a practical scheme that recommends stocks from S\&P 500 using machine learning. Our basic idea is to buy and hold the top 20\% stocks dynamically. First, we select representative stock indicators with good explanatory power. Secondly, we take five frequently used machine learning methods, including linear regression, ridge regression, stepwise regression, random forest and generalized boosted regression, to model stock indicators and quarterly log-return in a rolling window. Thirdly, we choose the model with the lowest Mean Square Error in each period to rank stocks. Finally, we test the selected stocks by conducting portfolio allocation methods such as equally weighted, mean-variance, and minimum-variance. Our empirical results show that the proposed scheme outperforms the long-only strategy on the S\&P 500 index in terms of Sharpe ratio and cumulative returns. This work is fully open-sourced at \href{https://github.com/AI4Finance-Foundation/Dynamic-Stock-Recommendation-Machine_Learning-Published-Paper-IEEE}{GitHub}.


\end{abstract}


\section{Introduction}
\label{sec:intro}


Earning reports play a key role in stock recommendation. Analysts use company earning reports to do stock buy-and-sell recommendation. Future earnings estimates are important factors to value a firm. Earnings forecasts are based on analysts' estimation of company growth and profitability. To predict earnings, most analysts build financial models that estimate prospective revenues and costs. However, it could be very difficult for analysts to accurately estimate earnings. Many researchers are trying to build a robust model to predict earnings. For example, earnings generated by the cross-sectional model are considered superior to analysts’ forecasts to estimate the implied costs of capital (ICC), which play a key role in firm valuation \cite{ICC2011}. Moreover, regression-based models \cite{luzhang2014} can be used to predict scaled and un-scaled net income \cite{regression2013}. Numerous recent papers consider using deep learning algorithms to model stock market data \cite{deeplearning1}. Deep neural networks models also can be trained to predict future fundamentals such as book-to-market ratio, and as a result, investors can use predicted fundamentals to rank current stocks \cite{deeplearning4}.


There are two traditional approaches. The first approach is selecting stocks based on a preset criteria such as price to earnings (P/E) ratio \cite{PE2012}. Stocks are ranked by P/E ratios using historical data. Then, a portfolio will contain stocks with lowest P/E ratios. This approach is unsatisfactory in practical situations since the selection with P/E ratio only is unstable (e.g., selecting top quantile stocks may result in a less predictive power.) The second approach jointly uses several criterion to rank stocks, such as P/E ratio, price to sales (P/S) ratio, price/earnings to growth (PEG) ratio, etc. However, this approach does not take the correlations among different predictor factors into consideration. Consequently, weights of these factors are assigned relatively subjective, which increases the risk.




Value investing has been widely used today by investors and portfolio managers. Graham first comes up with the concept of an “intrinsic” value for a stock that is independent of the market \cite{Graham1988}. He emphasizes the importance of an intrinsic value that is reflected by a company’s market size, assets level, dividends, financial strength, earnings stability,
earnings growth. Focusing on this value, he believes would prevent an investor from misjudgment and misinterpretation during a bull or bear market. In the long run, we expect stock prices should eventually be regression towards the company's intrinsic value. Many fundamental financial ratios such as P/E ratio, earnings per share (EPS), return on equity (ROE), profit margin and quick ratio indicate overall profitability, stability, operating efficiency, capital structure, ability of generating future cash flows and other valuable information of the corresponding companies. Thus, these financial ratios could reflect a company’s intrinsic value and should have predictive power on the future performance \cite{value1996}. Additionally, financial ratios provide normalization so that all companies would have the same scale of data and thus, those with large capital will have equal influence.





In this paper, we propose a novel scheme that predict stock's future price return based on earnings factors by machine learning. We use five machine learning algorithms (linear regression, random forest, ridge, stepwise regression, and generalized boosting regression) to assign weights to each factor dynamically, and select top $20\%$ stocks each quarter based on the ranking of predicted returns generated by the best performing algorithm over the past training periods before each re-balancing day on a rolling basis. The five models also have high explanatory power. By using the lowest MSE to choose the best model, we provide reliability to business decision, thus increase the security to financial investments. After we select stocks for our portfolio on each re-balancing day, we test asset allocation methodologies: mean-variance, min-variance, and equally-weighted allocation on selected stocks using in sample data (1990-2007). Risk management is involved by the method of maximum Sharpe ratio used in portfolio allocation methodologies. We are aiming to balance the expected return and the standard deviation of the portfolio to achieve the best risk to reward. Finally, we compare the P\&L of our strategy with S\&P 500 index \footnote{The Standard \& Poor's 500 is an American stock market index based on the market capitalizations of 500 large companies having common stock listed on the NYSE or NASDAQ.}, all three portfolio allocation methods outperform the market, summarize the competency of our strategy.

This paper proceeds as follows. Section II describes the rolling window, trading time, the data, and also presents the methodology and implementation of our scheme. Section III contains the portfolio allocation methods, risk management and transaction cost. Section IV presents the performance and Section V concludes the paper.

\section{Proposed Stock Recommendation Scheme}





\subsection{Rolling Window Based Data Separation}

Rolling windows can be employed to divide data for multiple purposes (i.e., training and testing). Rolling windows for training ranges from 16-quarter (4-year) to a maximum of 40-quarter (10-year). This training rolling window is followed by an one-year window for testing and we trade according to the test results. The training-testing-trading cycle of our strategy can be summarized by Fig. 1. We also extend the trade date by two months lag beyond the standard quarter end date in case some companies have a non-standard quarter end date, e.g. Apple released its earnings report on 2010/07/20 for the second quarter of year 2010. Thus for the quarter between 04/01 and 06/30, our trade date is adjusted to 09/01 (same method for other three quarters).

\begin{figure*}
    \centering
    \includegraphics[totalheight=5.5cm]{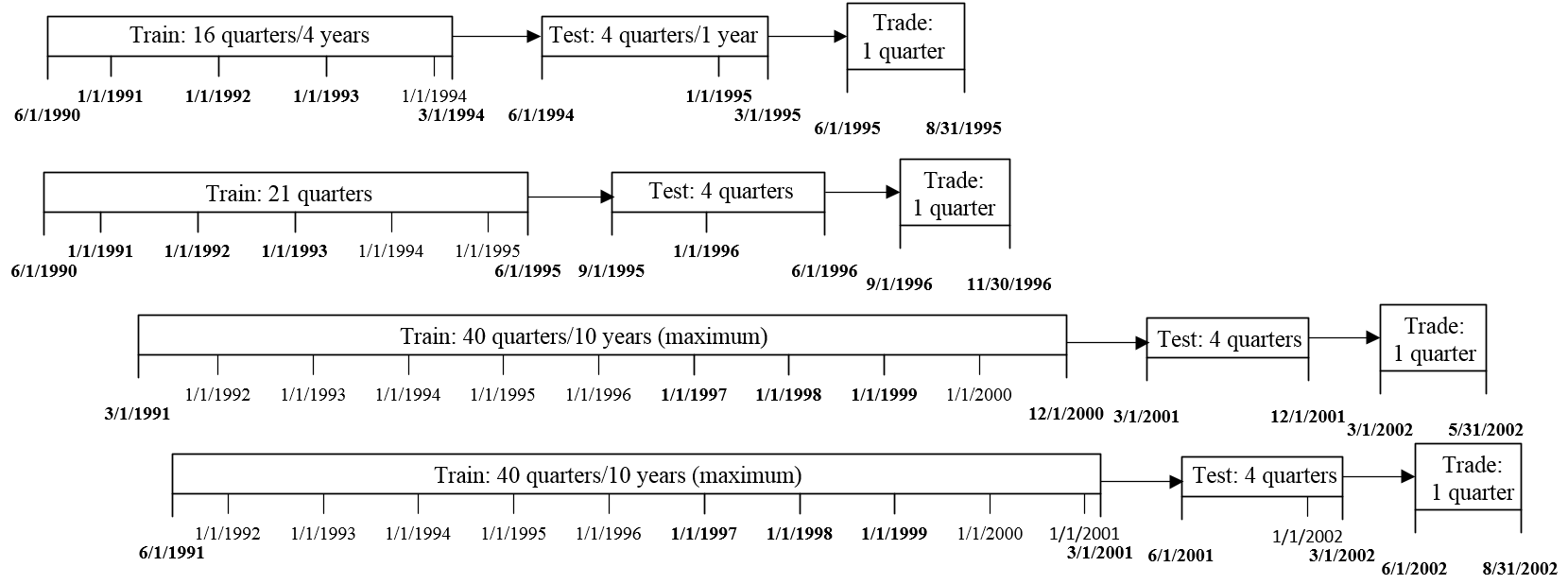}
    \caption{Rolling Window Based Data Seperation: Training rolling window is followed by a testing rolling window. There is a two-month delay after the end of the training rolling window.}
    \label{fig:}
\end{figure*}

\subsection{Data Preprocessing}

The data for this project is mainly taken from Compustat database accessed through Wharton Research Data Services (WRDS) \cite{wrds1}. The dataset used here consists of the data over the period of 27 years (from 06/01/1990 to 06/01/2017). We use all historical S\&P 500 component stocks (about 1142 stocks) as the S\&P 500 pool are updated quarterly. The adjusted close price goes on a daily basis (trading days) and generates 6,438,964 observations. The fundamental data goes on a quarterly basis and generates 91,216 observations. In addition, we delete outlier records that indicate a release date (rdq) after the trade date, which include about 0.84\% of the dataset. We assure that on our trade date, 99\% of the companies have their earnings reports ready to be used. In order to preserve an out-of-sample period sufficiently long for back-testing the relationship, the dataset has been divided into three periods in Fig. 2.

\begin{figure}
    \centering
    \includegraphics[totalheight=4.5cm]{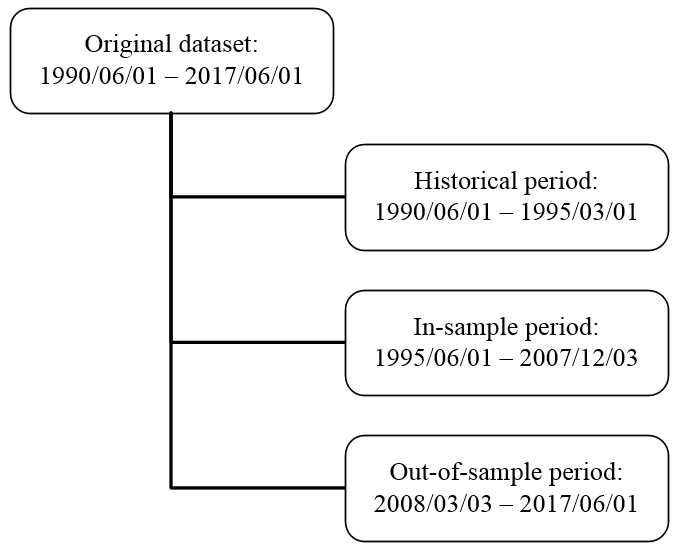}
    \caption{Dataset Division}
    \label{fig:}
\end{figure}

To build the dataset for training, we select top twenty most popular financial ratios in table I \cite{luzhang2014} and calculated these factors from the fundamental raw data from the WRDS. Also, in order to build a sector-neutral portfolio, we split the dataset by the Global Industry Classification Standard (GICS) sectors. We handle missing data separately by sector: if one factor has more than 5\% missing data, we delete this factor; if a certain stock generates the most missing data, we delete this stock. In this way, we've removed 46 stocks and the overall missing data is reduced to less than 7\% of each sector. Finally, we delete this 7\% missing data.

\begin{table}[t]
\centering
\caption{$20$ Financial Indicators}
 \scalebox{0.85}{ \begin{tabular}{|c|c|}
  \hline
      Revenue Growth &  Price to cash flow ratio  \\
  \hline
      Earnings per share (EPS) &Cash ratio\\
  \hline
      Return on asset (ROA) &Enterprise multiple  \\
  \hline
      Return on equity (ROE) &Enterprise value/cash flow from operations  \\
  \hline
      Price to earnings (P/E) ratio &Long term debt to total assets  \\
  \hline
      Price to sales (P/S) ratio &Working capital ratio  \\
  \hline
     Net profit margin &Debt to equity ratio \\
  \hline
      Gross profit margin  &Quick ratio\\
  \hline
      Operating margin &Days sales of inventory  \\
  \hline
      Price to book (P/B) ratio &Days payable of outstanding  \\
  \hline

  \end{tabular}}
\end{table}


\subsection{Methodology}

Our goal is to predict S\&P 500 forward quarter log-return $r^{qtr}_{T+f}$ given predictors $X_{T}$ constructed from historical data of the twenty financial factors over a particular quarter $T$ and S\&P 500 horizon $f$. At a given time $T$ of the financial horizon, the 1-quarter forward log-returns of a certain stock price $S$ are defined as:
\begin{equation}  
r^{qtr}_{T+f,i} =\ln(S_{T+f,i} / S_{T,i} ), \qquad  i=1,...,n_T, \end{equation}
where $n_T$ is the companies whose stock price and earnings factors are available at time T. 

A general estimator is the ordinary least square:
\begin{equation} r^{qtr}_{T+f,i}=\beta_0 + \sum_{j=1}^{p} \beta_jX_{T,i,j}+ \epsilon, \qquad j = 1,...,20, \end{equation} 
where $j$ is the number of the twenty financial ratios, $p$ is the total factors we used in the model, $\beta_0$ is the intercept of the model, $X_j$ corresponds to the $j$th predictor variable of the model, $\beta_j$ is the coefficients of the predictor variable and $\epsilon$ is the random error with expectation 0 and variance $\sigma^2$. Moreover, regularized linear OLS estimators have a higher accuracy in many aspects \cite{textbook1}. We need to use multiple regression estimators to increase accuracy. \cite{regression2013} has summarized the prediction rule and estimator selection rule for using multiple estimators:
\begin{equation} r^{qtr}_{T+f,i}|X_{T,i,j},\theta, \qquad  i=1,...n_T, \ j=1,...,20,\end{equation} 
\begin{equation} r^{qtr}_{t+f,i}=g_\theta(X_{t,i,j}) + \epsilon, 
 \quad t = T,...,T-h, \ i=1,...,n_t,\end{equation} 
where $h$ is the historical estimation period, $g_\theta(X_{t,i,j})$ is used to estimate $\theta$ through historical regressions. It is noticeable to point out that (2) is the basic estimator of (4). 

We pick five models for $g_\theta$: linear regression, forward and backward stepwise regression under Akaike information criterion (AIC), regularized linear OLS estimator ridge regression, tree based nonlinear model random forest and generalized boosted regression model (GBM) using gaussian distribution which implements AdaBoost algorithm and Friedman's gradient boosting machine. All of the algorithms are facilitated by standard R packages \cite{r2011}. For linear regression and stepwise regression we use $lm$ and $step$ and for ridge we use $glmnet$ and \emph{MASS} \cite{textbook2,leastsquare1}. For random forest we use $randomForest$ \cite{randomforest2002}. For gbm we use $gbm$ \cite{gbm2007}. The reason of using these five models is that we need feature selection methods to remove undesirable features, thus reducing the overfitting issues, improving model accuracy and expediting the training procedure. We also have a white-box model that we can observe every single factor with its coefficients in our model. 

Mean Squared Error (MSE) \cite{regression2013} is used as the metric for our evaluation.

\subsection{Implementation}


Our implementation can be summarized as the following four steps:

\textbf{Step 1}. Train and test the model to get the MSE for each of the five models. Our current methodology basically selects the minimum MSE. We assign $1$ to the selected model and $0$ to other models.

\textbf{Step 2}. Choose the model that has the lowest MSE in that certain period. For example, in Table II, we choose Ridge regression as our model to select stocks on Jun. 1$^{st}$, 1995. We choose Random Forest as our model to select stocks on Sept. 1$^{st}$, 1995.

\begin{table}[t]
\centering
\caption{Model Error and Selected Model for Sector 10, Energy}
 \scalebox{0.85}{ \begin{tabular}{|c|c|c|c|c|c|c|}
  \hline
    trading date & MSE linear & MSE RF & MSE ridge & MSE step &MSE gbm\\
  \hline
      19950601 & $0.02238$   &  $0.02180$  & $0.02161$   & $0.02205$   & $0.02443$    \\
  \hline
      19950901 & $0.01908$   &  $0.01828$  & $0.01870$   & $0.01841$   & $0.02098$    \\
  \hline
      19951201 & $0.01852$   &  $0.01641$  & $0.01820$   & $0.01855$   & $0.01996$    \\
  \hline
      19960301 & $0.02040$   &  $0.01822$  & $0.01981$   & $0.01879$   & $0.02192$    \\
  \hline
      19960603 & $0.02442$   &  $0.01885$  & $0.02394$   & $0.02340$   & $0.02210$    \\
  \hline
  \end{tabular}}
\end{table}

\textbf{Step 3}. Use the predicted return in the selected model to pick up top 20\% stocks from each sector.  We predict next quarter return (predicted y) using current information (test Xs) based on the trained model.

In this example, in Table III we use ridge predicted return to pick stocks for the trade period 1995/06/01, the selected top 20\% stocks are: WMB, OKE, RRC, PXD, VLO, EQT, HES, BHI, MUR, and NE. We then trade these stocks during the period between 1995/06/01 and 1995/09/01. In the second trade period of 1995/09/01, in Table IV we use random forest to pick the stocks, the top 20\% stocks are: CHK, SFS.1, WMB, RRC, VLO, BJS.1, MDR, PZE.1, HP, and CVX.  We then trade these stocks from 1995/09/01 to 1995/12/01. As for the stocks owned at previous quarters, such as WMB, RRC, and VLO, we just need to use the portfolio weights to adjust their shares.

\begin{table}[t]
\centering
\caption{Predicted Return on trade date: 1995/06/01 Sector 10, Energy}
 \scalebox{0.85}{ \begin{tabular}{|c|c|c|c|c|c|}
  \hline
    &Linear return & RF return & Ridge return & Step return & GBM return \\
  \hline
      WMB & $10.42\%$   &  $5.12\%$  & $9.24\%$   & $9.01\%$   & $3.51\%$    \\
  \hline
      OKE & $7.55\%$   &  $4.12\%$  & $7.42\%$   & $8.53\%$   & $2.56\%$    \\
  \hline
      RRC & $4.16\%$   &  $7.01\%$  & $3.74\%$   & $4.84\%$   & $1.83\%$    \\
  \hline
      PXD & $4.63\%$   &  $0.96\%$  & $3.66\%$   & $3.91\%$   & $0.23\%$    \\
  \hline
      VLO & $3.48\%$   &  $2.99\%$  & $3.47\%$   & $4.04\%$   & $2.56\%$    \\
  \hline
      EQT & $2.47\%$   &  $4.78\%$  & $2.34\%$   & $2.36\%$   & $1.83\%$    \\
  \hline
      HES & $1.80\%$   &  $4.30\%$  & $1.61\%$   & $1.81\%$   & $0.38\%$    \\
  \hline
      BHI & $1.33\%$   &  $-0.70\%$  & $1.15\%$   & $1.99\%$   & $-0.27\%$    \\
  \hline
      MUR & $1.11\%$   &  $1.07\%$  & $1.01\%$   & $0.49\%$   & $0.38\%$    \\
  \hline
      NE & $1.16\%$   &  $-6.33\%$  & $0.94\%$   & $0.85\%$   & $-2.21\%$    \\
  \hline
  \end{tabular}}
\end{table}

\begin{table}[t]
\centering
\caption{Predicted Return on trade date: 1995/09/01 Sector 10, Energy}
 \scalebox{0.85}{ \begin{tabular}{|c|c|c|c|c|c|}
  \hline
    &Linear return & RF return & Ridge return & Step return & GBM return \\
  \hline
      CHK & $0.97\%$    &  $12.45\%$  & $2.17\%$   & $4.86\%$   & $0.03\%$    \\
  \hline
      SFS.1 & $4.69\%$  &  $7.35\%$  & $4.27\%$   & $4.49\%$   & $0.78\%$    \\
  \hline
      WMB & $5.87\%$    &  $6.37\%$  & $5.15\%$   & $5.46\%$   & $0.77\%$    \\
  \hline
      RRC & $1.78\%$    &  $6.13\%$  & $1.52\%$   & $1.76\%$   & $0.09\%$    \\
  \hline
      VLO & $2.50\%$    &  $4.83\%$  & $2.65\%$   & $3.01\%$   & $0.77\%$    \\
  \hline
      BJS.1 & $5.36\%$  &  $4.23\%$  & $5.28\%$   & $6.38\%$   & $-1.71\%$    \\
  \hline
      MDR & $1.54\%$    &  $3.96\%$  & $1.26\%$   & $1.14\%$   & $0.77\%$    \\
  \hline
      PZE.1 & $0.96\%$  &  $3.78\%$  & $0.55\%$   & $0.78\%$   & $-1.71\%$    \\
  \hline
      HP & $-1.63\%$    &  $3.50\%$  & $-1.44\%$   & $-1.62\%$   & $0.77\%$    \\
  \hline
      CVX & $-0.73\%$   &  $3.19\%$  & $-0.71\%$   & $-1.12\%$   & $0.09\%$    \\
  \hline
  \end{tabular}}
\end{table}

\textbf{Step 4}. We check on the corresponding models’ features and its coefficients or importance level to ensure that there are no abnormal results. in Table V and VI, e.g. assign 0 to all features.

\begin{table}[t]
\centering
\caption{Ridge Coefficients: 1995/06/01}
 \scalebox{0.85}{ \begin{tabular}{|c|c|c|c|}
  \hline
    Factor & Coefficient & Factor & Coefficient \\
  \hline
      ROA &     $0.205677$  & DPO &     $0.000004$ \\
  \hline
      GPM &     $0.116841$ & DSI &     $-0.000018$\\
  \hline
      REVGH &     $0.081318$ &  EM &     $-0.000405$\\
  \hline
      (Intercept) &     $0.049237$ &  WCR &     $-0.000422$\\
  \hline
      NPM &     $0.015640$  &PS &     $-0.002042$\\
  \hline
      CR &     $0.004123$ & PCFO &     $-0.003140$   \\
  \hline
      EPS &     $0.002562$ &LTDTA &     $-0.026227$\\
  \hline
      QR &     $0.002322$ & PB &     $-0.032136$\\
  \hline
      DE &     $0.000612$ & OM &     $-0.055619$  \\
  \hline
      EVCFO &     $0.000013$ & ROE &     $-0.078489$ \\
  \hline
  PE &     $0.000004$  &&\\
  \hline
  \end{tabular}}
\end{table}

\begin{table}[t]
\centering
\caption{Random Forest Importance Table: 1995/09/01}
 \scalebox{0.85}{ \begin{tabular}{|c|c|c|c|}
  \hline
    Factor & Importance & Factor & Importance \\
  \hline
      PB &     $14.2818$  &PE &     $5.6404$\\
  \hline
      EPS &     $9.0556$ & CR &     $5.5080$ \\
  \hline
      ROE &     $8.6929$  &ROA &     $5.3972$\\
  \hline
      PS &     $8.4976$ &PCFO &     $5.1313$ \\
  \hline
      WCR &     $8.4904$  &EVCFO &     $4.6672$ \\
  \hline
      EM &     $7.3808$  & LTDTA &     $4.6447$ \\
  \hline
      GPM &     $7.2091$  & OM &     $4.2249$   \\
  \hline
      QR &     $7.1337$  &DE &     $3.3296$ \\
  \hline
      DPO &     $6.6917$ & DSI &     $2.4621$   \\
  \hline
      NPM &     $5.9743$ &REVGH &     $0.1632$   \\

  \hline
  \end{tabular}}
\end{table}

We finish these steps for all eleven GICS sectors. Then we get a final table of all selected stocks with its tic name, predicted returns for next quarter, and the corresponding trading periods to conduct portfolio allocation.

\section{Portfolio Allocation and Risk Management}


Portfolio allocation is crucial to an investment strategy because its balance risk and return by modeling individual asset's weights. Mean-variance and minimumu-variance are two typical methods for portfolio allocation. They perform diversification by constraining mean, volatility and correlation inputs to reduce sampling error \cite{meanvariance2012}. In our portfolio we used mean-variance and min-variance to decide the weights of each stock, and then use equal-weighted portfolio as our benchmark. We perform these methods by Matlab Financial Toolbox-Portfolio Object \cite{matlab2017}.

\subsection{Mean-Variance and Minimum-Variance Constraints}

We first use mean-variance optimization to allocate the stocks we have picked. In. Fig. 3. the yellow star on the curve is the mean-variance result during our first trade time; the rest of the points are stocks plotted based on its predicted return and standard deviation. The result shows that our approach is legitimate. 

We set the following constraints for mean-variance:

$\bullet$  Expected return: predicted return of next quarter.

$\bullet$  Covariance matrix: use 1 year historical daily return.

$\bullet$  Long only: upper bound 5\% and Lower bound 0\%.

$\bullet$  Fully invest our capital: sum of weights=100\%.

$\bullet$  Take no leverage: LowerBudget = UpperBudget = 1.

\begin{figure}
    \centering
    \includegraphics[width=85mm]{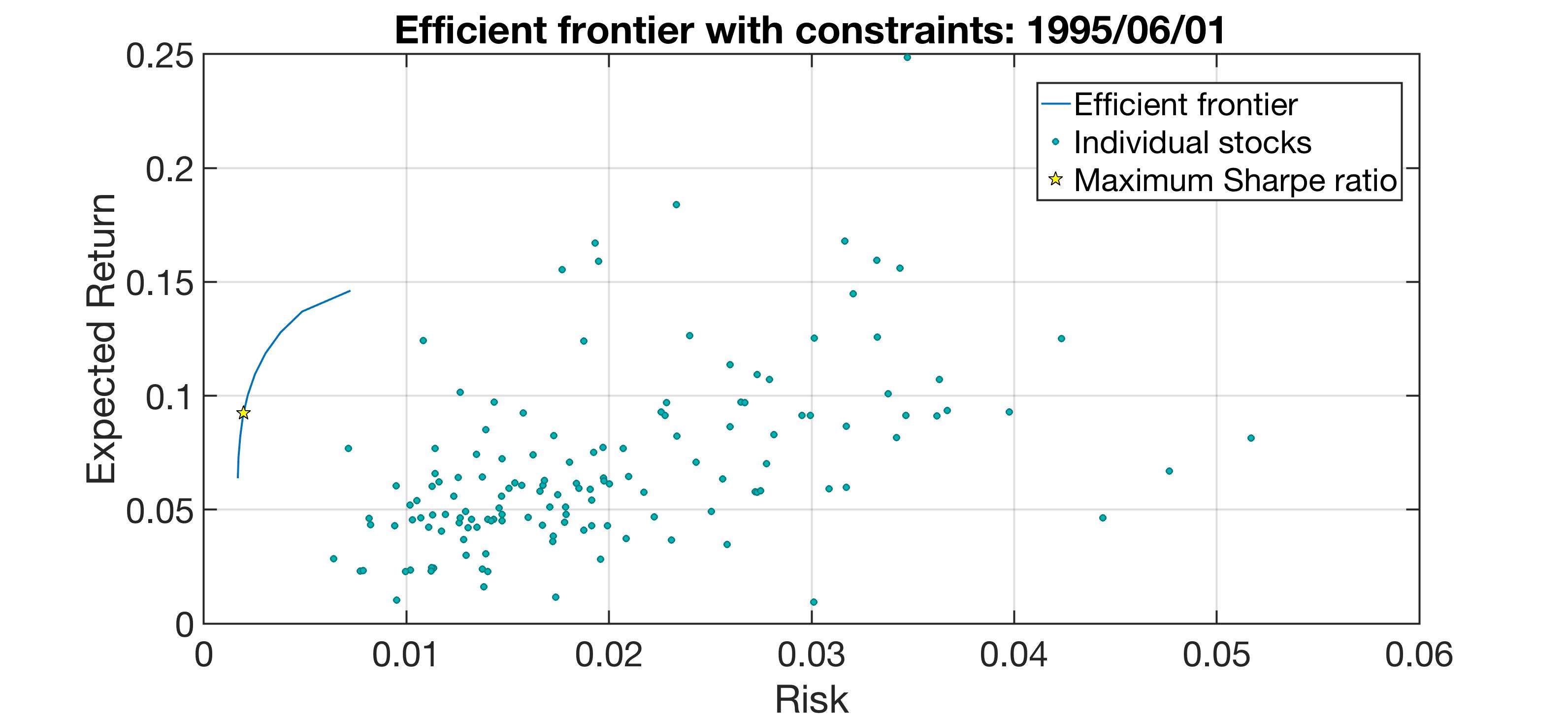}
    \caption{This figure shows the efficient frontier.}
    \label{fig:}
\end{figure}

Then, we try the min-variance optimization approach to allocate the stocks we have picked. The criteria are almost the same as the mean-variance method except that we set the expected return to be 0.

\subsection{Transaction Costs}

Generally, fees for each trade are measured based on broker fees, exchange fees and SEC fees. In the real-world scenarios, a fund or trading firm might have different execution costs for many reasons. Despite these possible variations in cost, after going through several scenarios we consider our transaction cost to be 1/1000 of the value of that trade. We believe our fee assumption to be sufficient and reasonable for the study. 

We use the following formula to calculate the transaction cost: 
\begin{equation}  
\sum_{i=1}^{n}|S_{t,i}-S_{t-1,i}|\cdot P_i \times 0.1\%
\end{equation}
where $S_{t,i}$ is the shares we need to buy or sell share based on the portfolio weights at current time $t$ and $S_{t-1,i}$ is the shares left at previous time $t-1$. $P_i$ is the current stock price of stock $i$.

\subsection{Risk Management}
After the procedure of building portfolio and structuring with appropriate functions, we equip decision rules that would be applied to risk management of each trade. Fundamentally, due to the nature of long-only strategy, the risk was controlled internally through our portfolio optimization methods. We minimize variance and maximum Sharpe ratio, have limits on position sizes (maximum of position size is 5\% of portfolio value), and don't take any leverage.

\section{Performance Evaluation}

\begin{figure}
    \centering
    \includegraphics[width=85mm]{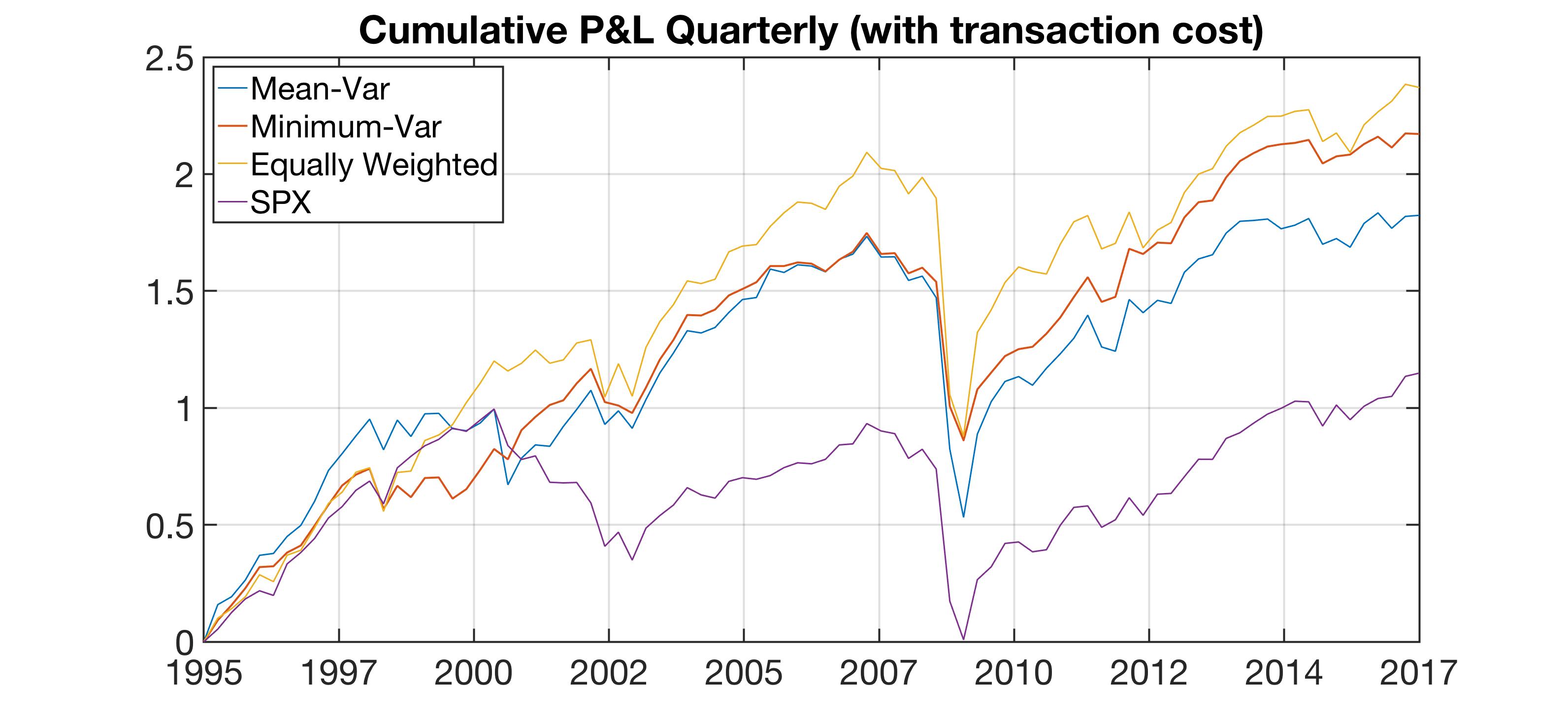}
    \caption{This figure shows the P\&L.}
    \label{fig:}
\end{figure}

\begin{figure}
    \centering
    \includegraphics[width=85mm]{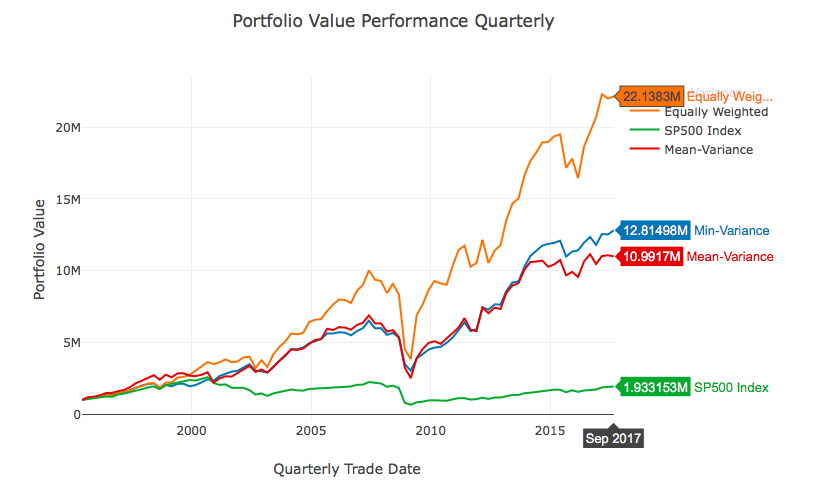}
    \caption{This figure shows Portfolio Value starts with 1 million.}
    \label{fig:}
\end{figure}

In Fig. 4, the results show that our strategy outperforms the market. The following back-test figures including equally weighted portfolio, mean-variance portfolio and minimum-variance portfolio indicate better performance than that of the benchmark S\&P 500 index. More importantly, all the statistics can show that the portfolio outperforms the S\&P 500 index not only in the in-sample training period (before 2007) but also in the overall trade period in Table VIII. Our codes are available online \cite{codes}.

We conclude that our scheme for stock selections does generate a better result than the market portfolio does. If we only analyze the portfolio value performance from the figure, it is noticeable that the equally weighted portfolio, the benchmark, has higher value than min-variance and mean-variance portfolio. However, the portfolio value is not the only consideration when selecting the optimal portfolio. We have two reasons to conclude that the min-variance portfolio is a better method in the real trade period. First, the equally-weighted portfolio is not robust enough. We notice that the performance of this method fully depends on the predicted returns that we calculate from our model. The predicted returns will vary every time we run our model. Secondly, min-variance portfolio allocation takes the risk factor into consideration, making it more reliable in real trade. From the Table VII, we find that the min-variance allocation has a higher Sharpe ratio than that of the equally-weighted allocation during the in-sample period. We thus choose the min-variance as our portfolio allocation method.

\begin{table}[t]
\centering
\caption{In sample data result: 1995-2007}
 \scalebox{0.95}{ \begin{tabular}{|c|c|c|c|c|}
  \hline
   & Mean-Var & Equally & Min-Var & S\&P 500\\
   \hline
  Annualized Return & $13.17\%$ & $16.12\%$ & $13.29\%$ & $7.12\%$       \\
  \hline
  Annualized Std    & $17.0\%$ & $16.4\%$ & $12.9\%$ & $13.8\%$      \\
  \hline
  Sharpe Ratio     & $0.687$   &  $0.887$  & $0.917$   & $0.406$      \\
  \hline
  \end{tabular}}
\end{table}

\begin{table}[t]
\centering
\caption{Overall Performance}
 \scalebox{0.85}{ \begin{tabular}{|c|c|c|c|c|}
  \hline
   (Risk-Free: 1.5\%) & Mean-Var & Equally & Min-Var & S\&P 500\\
   \hline
    Start Value in million    & $1$   &  $1$  & $1$   & $1$      \\
  \hline
    End Value in million    & $10.9917$   &  $22.1383$  & $12.81498$   & $1.933153$      \\
  \hline
    Total Return    & $999.17\%$   &  $2113.83\%$  & $1181.50\%$   & $93.32\%$      \\
  \hline
      Maxiumum Drawdown    & $-56.89\%$   &  $-57.63\%$  & $-46.30\%$   & $-66.73\%$      \\
  \hline
   Annualized Return & $8.29\%$ & $10.77\%$ & $9.87\%$ & $5.22\%$       \\
  \hline
  Annualized Std    & $23.6\%$ & $26.4\%$ & $18.1\%$ & $19.1\%$      \\
  \hline
  Sharpe Ratio     & $0.287$   &  $0.351$  & $0.462$   & $0.195$      \\
  \hline
  \end{tabular}}
\end{table}

\section{Conclusion}
Applying machine learning algorithms to the fundamental financial data can filter out stocks with a relative bad earnings, thus providing a better way to select stocks. Minimum-variance method, the 5 percent holding rule, no short and leverage rule provide risk management and diversification, reduce the portfolio risk and thus yield a higher Sharpe ratio. Compared to the benchmark, our trading strategy outperforms the S\&P 500 index. More importantly, combined with our trading strategy, the portfolio allocation method is proven to improve the overall performance. Finally, the Sharpe ratios of the three portfolio methods indicate that our strategy also outperforms the market.  Future work would be dealing with anomaly data \cite{liu2017ls} in the data preprocessing stage, and applying accurate prediction schemes by modeling stock indicators as tensor time series \cite{liu2017fourth} with sparsity in transform domains.

\bibliographystyle{IEEEbib}
\bibliography{reference}

\end{document}